\documentclass{article}
\usepackage{graphicx}
\usepackage{amssymb,amsmath}
\usepackage{multirow}
\date{}

\makeatletter \@addtoreset{equation}{section}
\renewcommand{\theequation}{\thesection.\@arabic\c@equation}
\newcommand{\affiliation}[1]{\let\thefootnote\relax\footnote{\mbox{}\\ \noindent {#1}}}

\makeatother

\begin{document}
\title{\textbf{A New Recursive Algorithm For Inverting A General Comrade Matrix}}

\author{ A. A. KARAWIA\footnote{ Home Address: Mathematics Department, Faculty of Science, Mansoura University,
Mansoura 35516, Egypt. E-mail:abibka@mans.edu.eg}\\
Computer science unit, Deanship of educational services, Qassim University,\\
 P.O.Box 6595, Buraidah 51452, Saudi Arabia. \\
              E-mail: kraoieh@qu.edu.sa
}

\maketitle
%\large{
\begin{abstract}
In this paper, the author present a reliable symbolic computational algorithm for inverting a general comrade matrix by using parallel computing along with recursion. The computational cost of our algorithm is $O(n^2)$. The algorithm is implementable to the Computer Algebra System (CAS) such as MAPLE, MATLAB and MATHEMATICA. Three examples are presented for the sake of illustration.\bigskip
\end{abstract}

\begin{flushleft}\footnotesize
\hspace*{0.9cm}{\textbf{Keywords}:Comrade matrices; LU factorization; Inverse matrix; Computer algebra systems(CAS). \\
\hspace*{0.9cm}{\textbf{AMS Subject Classification}:15A15; 15A23; 68W30; 11Y05; 33F10; F.2.1; G.1.0.\\}
}

\end{flushleft}
\newtheorem{alg}{Algorithm}[section]

\section{Introduction}

\hspace*{0.5cm} The $n\times n$ general comrade matrix, denoted by $C$, takes the form
\begin{equation}
C=\left(
           \begin{array}{ccccccc}
             \beta_1 & \alpha_1 & 0 & \cdots & \cdots& & 0 \\
             \gamma_2 & \beta_2 & \alpha_2 & \ddots & & & 0 \\
             0 & \gamma_3 & \beta_3 & \alpha_3 &\ddots & & 0\\
             \vdots & \ddots & \ddots & \ddots & \ddots& \ddots& 0 \\
             \vdots &  & \ddots & \ddots & \ddots&\ddots & 0 \\
             0 & \cdots & \cdots &0 & \gamma_{n-1} & \beta_{n-1}  & \alpha_{n-1} \\
             a_n & a_{n-1} & \cdots&a_4 & a_3 & \gamma_n & \beta_n \\
           \end{array}
         \right)
,\quad n\ge3.
\end{equation}
\\
The comrade matrix is a generalization of the companion matrix and is associated with a polynomial expressed as a linear combination of an arbitrary orthogonal basis. This matrix appears frequently in many areas of science and engineering, for example in linear multivariable systems theory[1], Computing the Greatest Common Divisor of Polynomials[2] and division of generalized polynomials [3]. The solution of comrade linear system has been investigated by many authors (see for instance, [4-6]). Finding the inverse of comrade matrix is usually required to solve this linear system. The motivation of the current paper is to establish efficient algorithm for inverting companion matrix of the form (1.1).\\

The paper is organized as follows. In Section 2, new symbolic computational algorithm,that will not break, is constructed. In Section 3, three illustrative examples are given. Conclusions of the work are presented in Section 4.

\section{Main Results}

In this section we shall focus on the construction of new symbolic computational algorithm for computing the determinant and the inverse of a general comrade matrix. Firstly, we begin with computing the $LU$ factorization of the matrix $C$. It is as in the following:

Let
\begin{equation}
C=LU
\end{equation}
where
\begin{equation}
L=\left(
           \begin{array}{ccccccc}
             1 & 0 & 0 & \cdots & \cdots&\cdots & 0 \\
             \frac{\gamma_2}{\mu_1} & 1 & 0 & \ddots & & & 0 \\
             0 & \frac{\gamma_3}{\mu_2} & 1 & 0 &\ddots & & 0\\
             \vdots & \ddots & \ddots & \ddots & \ddots& \ddots& 0 \\
             \vdots &  & \ddots & \ddots & \ddots&\ddots & 0 \\
             0 & \cdots & \cdots &0 & \frac{\gamma_{n-1}}{\mu_{n-2}} & 1  & 0 \\
             x_1 & x_2 & \cdots& \cdots & x_{n-2} & x_{n-1} & 1 \\
           \end{array}
         \right)
\end{equation}
and
\begin{equation}
U=\left(
           \begin{array}{ccccccc}
             \mu_1 & \alpha_1 & 0 & \cdots & \cdots&\cdots & 0 \\
             0 & \mu_2 & \alpha_2 & \ddots & & & 0 \\
             0 & 0 & \mu_3 & \alpha_3 &\ddots & & 0\\
             \vdots & \ddots & \ddots & \ddots & \ddots& \ddots& 0 \\
             \vdots &  & \ddots & \ddots & \ddots&\ddots & 0 \\
             0 & \cdots & \cdots &0 & 0 & \mu_{n-1}  & \alpha_{n-1} \\
             0 & 0 & \cdots& \cdots & 0 & 0 & \mu_n \\
           \end{array}
         \right)
\end{equation}
The elements in the matrices L and U in (2.2) and (2.3) satisfy:

\begin{equation}
\mu_i=\left\{\begin{matrix}
\beta_1  &\text{if}\quad i=1 \\
 \beta_i-\frac{\alpha_{i-1}}{\mu_{i-1}}\gamma_i &\quad\quad\quad\quad\quad\text{if}\quad i=2,3, ..., n-1 \\
 \beta_n-\alpha_{n-1}x_{n-1} &\text{if}\quad i=n,
\end{matrix}\right.
\end{equation}
and
\begin{equation}
x_i=\left\{\begin{matrix}
\frac{a_n}{\mu_1}  &\text{if}\quad i=1 \\
 \frac{1}{\mu_i}(a_{n-i+1}-\alpha_{i-1}x_{i-1}) &\quad\quad\quad\quad\quad\text{if}\quad i=2,3, ..., n-2 \\
 \frac{1}{\mu_{n-1}}(\gamma_n-\alpha_{n-2}x_{n-2}) &\quad\text{if}\quad i=n-1.
\end{matrix}\right.
\end{equation}

We also have:
\begin{equation}
Det(C)=\prod_{i=1}^n\mu_i.
\end{equation}
At this point it is convenient to formulate our first result. It is a symbolic algorithm for
computing the determinant of a comrade matrix $C$ of the form (1.1).

\begin{alg}
To compute Det($C$) for the comrade matrix $C$ in (1.1), we may proceed as follows: \\
\textbf{INPUT} order of the matrix $n$ and the components $\alpha_i$, $i=1, 2, ..., n-1$, $\beta_i$, $i = 1, 2, . . . , n$, \\
\hspace*{1.5cm} $\gamma_i$, $i = 2, 3, . . . , n$, and $a_i$, $i=3, 4, ..., n$.\\
\textbf{OUTPUT} The determinant of comrade matrix $C$.\\
\textbf{Step 1:} Set $\mu_1=\beta_1$. If $\mu_1= 0$ then $\mu_1= t$($t$ is just a symbolic name) end if. Set $x_1=\frac{a_n}{\mu_1}$.\\
\textbf{Step 2:} For $i=2, 3, ..., n-2$\\
   \hspace*{2.5cm}Compute $\mu_i = \beta_i-\frac{\gamma_{i}}{\mu_{i-1}}\alpha_{i-1}$, if $\mu_i= 0$ then $\mu_i= t$,\\
   \hspace*{2.5cm}Compute $x_i = \frac{1}{\mu_i}(a_{n-i+1}-\alpha_{i-1}x_{i-1})$,\\
 \textbf{Step 3:} Set $\mu_{n-1} = \beta_{n-1}-\frac{\gamma_{n-1}}{\mu_{n-2}}\alpha_{n-2}$, if $\mu_{n-1}= 0$ then $\mu_{n-1}= t$ end if,\\
   \hspace*{1.4cm}Set $x_{n-1}=\frac{1}{\mu_{n-1}}(\gamma_n-x_{n-2}\alpha_{n-2})$ .\\
   \hspace*{1.4cm}Set $\mu_n=\beta_n-\alpha_{n-1}x_{n-1}$, if $\mu_n= 0$ then $\mu_n= t$ end if.\\
\textbf{Step 4:} Compute $Det(C) =\Big( \prod_{i=1}^n\mu_i\Big)_{t=0}$.
\end{alg}

The symbolic Algorithm 2.1 will be referred to as \textbf{DETSGCM} algorithm. The computational cost of \textbf{DETSGCM} algorithm is $7n-10$ operations. The new algorithm \textbf{DETSGCM} is very useful to check the nonsingularity of the matrix $C$\\

Now, when the matrix $C$ is nonsingular, its inversion is computed as follows:\\
Let
\begin{equation}
C^{-1}=[S_{i,j}]_{1\le i,j\le n}=[Col_1, Col_2, ..., Col_n].
\end{equation}
where $Col_m$ denotes $m_{th}$ column of $C^{-1}$, $m = 1, 2, ...,n$.\\
Since the Doolittle LU factorization of the matrix $C$ in (1.1) is always possible then we can use parallel computations to get the elements of the last two columns $Col_i =(S_{1,i}, S_{2,i}, ..., S_{n,i})^T$ , $i = n$ and $n - 1$ of $C^{-1}$ as follows [7]:\\
Solving in parallel the standard linear systems whose coefficient matrix $L$ is given by (2.2)
\begin{equation}
L\left(
   \begin{array}{cc}
     Q_1^{(n)} & Q_1^{(n-1)} \\
     Q_2^{(n)} & Q_1^{(n-1)} \\
     \vdots & \vdots \\
      &  \\
     Q_{n-2}^{(n)} & Q_{n-2}^{(n-1)} \\
     Q_{n-1}^{(n)} & Q_{n-1}^{(n-1)} \\
     Q_n^{(n)} & Q_n^{(n-1)} \\
   \end{array}
 \right)=\left(
   \begin{array}{cc}
     0 & 0 \\
     0 & 0 \\
     \vdots & \vdots \\
      &  \\
     0 & 0 \\
     0 & 1 \\
     1 & 0 \\
   \end{array}
 \right)
\end{equation}
we get
\begin{equation}
\left(
   \begin{array}{cc}
     Q_1^{(n)} & Q_1^{(n-1)} \\
     Q_2^{(n)} & Q_1^{(n-1)} \\
     \vdots & \vdots \\
      &  \\
     Q_{n-2}^{(n)} & Q_{n-2}^{(n-1)} \\
     Q_{n-1}^{(n)} & Q_{n-1}^{(n-1)} \\
     Q_n^{(n)} & Q_n^{(n-1)} \\
   \end{array}
 \right)=\left(
   \begin{array}{cc}
     0 & 0 \\
     0 & 0 \\
     \vdots & \vdots \\
      &  \\
     0 & 0 \\
     0 & 1 \\
     1 & -x_{n-1} \\
   \end{array}
 \right)
\end{equation}
Hence, solving the following standard linear systems whose coefficient matrix $U$ is given by (2.3)
\begin{equation}
U\left(
   \begin{array}{cc}
     S_{1,n} & S_{1,n-1} \\
     S_{2,n} & S_{2,n-1} \\
     \vdots & \vdots \\
      &  \\
     S_{n-2,n} & S_{n-2,n-1} \\
     S_{n-1,n} & S_{n-1,n-1} \\
     S_{n,n} & S_{n,n-1} \\
   \end{array}
 \right)=\left(
   \begin{array}{cc}
     0 & 0 \\
     0 & 0 \\
     \vdots & \vdots \\
      &  \\
     0 & 0 \\
     0 & 1 \\
     1 & -x_{n-1} \\
   \end{array}
 \right)
\end{equation}
gives the two columns $Col_i$, $i = n$ and $n - 1$ in the forms:
\begin{equation}
S_{n,n}=\frac{1}{\mu_n},
\end{equation}

\begin{equation}
S_{i,n}=-\frac{\alpha_i S_{i+1,n}}{\mu_i},\quad i=n-1, n-2, ...,1,
\end{equation}

\begin{equation}
S_{n,n-1}=-\frac{x_{n-1}}{\mu_n},
\end{equation}

\begin{equation}
S_{n-1,n-1}=\frac{1-\alpha_{n-1}S_{n,n-1}}{\mu_{n-1}},
\end{equation}

\begin{equation}
S_{i,n-1}=\frac{-\alpha_iS_{i+1,n-1}}{\mu_i},\quad i=n-2, n-3, ...,1.
\end{equation}

Using equations (2.11)-(2.15) with the fact $C^{-1}C = I_n$ where $I_n$ is the $n\times n$ identity matrix, the elements in the remaining (n - 2) columns of $C^{-1}$ may be obtained recursively using are obtained by using

\begin{equation}
Col_{n-2}=\frac{1}{\alpha_{n-2}}(E_{n-1}-\beta_{n-1}Col_{n-1}-\gamma_nCol_n),
\end{equation}

\begin{equation}
Col_j=\frac{1}{\alpha_j}(E_{j+1}-\beta_{j+1}Col_{j+1}-\gamma_{j+2}Col_{j+2}-a_{n-j}Col_n),\quad j=n-3, n-4, ...,1.
\end{equation}

Here $E_r=(\delta_{1r}, \delta_{2r}, ..., \delta_{nr})^T,\quad r=1, 2, ..., n$, where $\delta_{ir}$ is the Kronecker symbol.\\

Now we formulate a second result. It is a symbolic computational algorithm to compute the inverse of a general comrade matrix of the form (1.1) when it exists.\\
\begin{alg}
To find the $n\times n$ inverse matrix of the general comrade matrix $C$ in (1.1) by using the relations (2.11)-(2.17). \\
\\
\textbf{INPUT} Order of the matrix $n$ and the components $\alpha_i$, $i=1, 2, ..., n-1$, $\beta_i$, $i = 1, 2, . . . , n$,\\
\hspace*{1.5cm} $\gamma_i$, $i = 2, 3, . . . , n$, and $a_i$, $i=3, 4, ..., n$.\\
\textbf{OUTPUT} Inverse comrade matrix $C^{-1}$.\\
\textbf{Step 1:} Use the \textbf{DETSGCM} algorithm to check the nonsingularity of the matrix $C$. If the matrix $C$ is singular \\
\hspace*{1.3cm} then OUTPUT ('The matrix $C$ is singular'), Stop.\\
\textbf{Step 2:} If $\mu_i = 0$ for any $i = 1, 2, ..., n$, set $\mu_i = t$ (t is just a symbolic name).\\
\textbf{Step 3:} If $\alpha_i = 0$ for any $i = 1, 2, ..., n-2$, set $\alpha_i = t$.\\
\textbf{Step 4:} For $i = 1, 2, ..., n$, compute and simplify the components $S_{i,n}$ and $S_{i, n-1}$ of the columns $Col_j$, \\
\hspace*{1.5cm}$j = n$, and $ n-1$, respectively, by using (2.11)-(2.15).\\
\textbf{Step 5:} For $i = 1, 2, ..., n$, compute and simplify the components $S_{i,n-2}$ by using (2.16).\\
\textbf{Step 6:} For $j = n - 3, n-4, ..., 1$, do\\
\hspace*{1.8cm} For $i = 1, 2, ..., n$, do\\
\hspace*{2.6cm} Compute and simplify the components $S_{i,j}$ by using (2.17).\\
\hspace*{1.8cm} End do\\
\hspace*{1.5cm} End do\\
\textbf{Step 7:} Substitute the actual value $t = 0$ in all expressions to obtain the elements,
$S_{i,j}$, $i, j = 1, 2, ..., n.$
\end{alg}

The symbolic Algorithm 2.2 will be referred to as \textbf{SGCMINV}. The computational cost of \textbf{SGCMINV} algorithm is $7n^2-5n-11$ operations. In[8], the author presented recurrence relations for the rows of an inverse comrade matrix but he supposed that $\alpha_i\ne0$ for $i=1,2, ...,n-1$ and the first row of an inverse comrade matrix is known. The computational cost of this method is $O(n^3)$ operations. On the other hand, if we set $a_i=0,\quad i=3,4, ...,n$, the Algorithm 2.3 in [7] will be special case of the \textbf{SGCMINV} algorithm.\\

\section{ILLUSTRATIVE EXAMPLES}
       In this section we give three examples for the sake of illustration.\\
\\
\textbf{Example 3.1.} Consider the $7\times 7$ matrices C given by\\ \\
\hspace*{2.5cm}$
C=\left(\begin{array}{ccccc}
                    -\frac{1}{2} & \frac{1}{2} & 0 & 0 & 0  \\
                    \frac{3}{5} & -\frac{4}{5} & \frac{1}{5} & 0 & 0  \\
                    0 & \frac{1}{3} & -\frac{2}{3} & \frac{1}{3} & 0  \\
                    0 & 0 & \frac{4}{2} & -\frac{5}{2} & \frac{1}{2} \\
                    -\frac{1}{3} & -\frac{1}{3} & -\frac{1}{3} & \frac{2}{3} & -\frac{3}{3}  \\
                    \end{array}
                \right)\\ \\
$
by applying the \textbf{SGCMINV} algorithm.\\
\begin{itemize}
  \item $\mu=(-\frac{1}{2}, -\frac{1}{5},-\frac{1}{3},-\frac{1}{2},-\frac{4}{3})$,\\
        $Det(C)=-\frac{1}{45}$(Step 1).\\
  \item $C^{-1}=\left(
                  \begin{array}{ccccc}
                    -24 & -\frac{75}{4}  & -\frac{39}{4} &-\frac{3}{2}& -34 \\
                    -22 & -\frac{75}{4}  & -\frac{39}{4} & -\frac{3}{2}&-34 \\
                    -16 & -\frac{55}{4}  & -\frac{39}{4} & -\frac{3}{2}&-34 \\
                    -10 & -\frac{35}{4}  & -\frac{27}{4} & -\frac{3}{2}&-34 \\
                    14 & \frac{45}{4}   & \frac{21}{4} & \frac{1}{2}&-34 \\
                  \end{array}
                \right)$ (Steps 2-7).
\end{itemize}
\textbf{Example 3.2.} Consider the $4\times 4$ matrices C given by\\ \\
\hspace*{2.5cm}$
C=\left(
  \begin{array}{cccc}
    0 & 1 & 0 & 0  \\
    2 & -1 & 5 & 0  \\
    0 & 3 & 1 & 2 \\
    -1 & 1 & 5 & 3
  \end{array}
\right)\\ \\
$
by applying the \textbf{SGCMINV} algorithm.\\
\\
\begin{itemize}
  \item $\mu=( t, -\frac{t+2}{t}, \frac{2(8t+1)}{t+2},\frac{2(7t-6)}{8t+1})$,\\
  $Det(C)=\Big[-4(7t-6)\Big]_{t=0}=24$(Step 1).\\
  \item $C^{-1}=\left(
  \begin{array}{cccc}
  \frac{7}{7t-6}& -\frac{7}{28t-24}& -\frac{15}{28t-24}& \frac{5}{14t-12} \\
  -\frac{6}{7t-6}& \frac{7t}{4(7t-6)}& \frac{15t}{4(7t-6)}& -\frac{5t}{2(7t-6)}\\
   -\frac{4}{7t-6}&\frac{7t-2}{4(7t-6)}&\frac{3(t+2)}{4(7t-6)}& -\frac{t+2}{2(7t-6)}\\
  \frac{11}{7t-6}& -\frac{14t-1}{4(7t-6)}& -\frac{5(2t+3)}{4(7t-6)}&\frac{8t+1}{2(7t-6)}
    \end{array}
\right)_{t=0}=\left(
  \begin{array}{cccc}
  -\frac{7}{6}& \frac{7}{24}& \frac{3}{8}& -\frac{5}{12} \\
  1& 0& 0& 0\\
   \frac{2}{3}&\frac{1}{12}&-\frac{1}{4}& \frac{1}{6}\\
  -\frac{11}{6}& -\frac{1}{24}& \frac{5}{8}&-\frac{1}{12}
    \end{array}
\right)$ (Steps 2-7).\\

\end{itemize}
\textbf{Example 3.3.} We consider the following $n\times n$ comrade matrix in order to demonstrate the efficiency of \textbf{SGCMINV} algorithm.\\
\\
$
C=\left(
  \begin{array}{cccccc}
    -3/2 & 1/2 & 0 & \cdots & \cdots & 0 \\
    1/2 & -3/2 & 1/2 & 0 & \ddots    & 0 \\
    0 & 1/2 & -3/2 & 1/2 & \ddots     & \vdots \\
    \vdots & \ddots & \ddots & \ddots & \ddots   & 0 \\
     0 &   \cdots & 0 & 1/2 & -3/2 & 1/2 \\
    -1/2 & -1/2 &  \cdots & -1/2 & (1-1)/2 & (-3-1)/2 \\
  \end{array}
\right)\\
$

We used \textbf{SGCMINV} algorithm to compute the inverse of comrade matrix $C$. Results are given in the next table in which $\varepsilon=||C^{-1}_\textbf{exact}-C^{-1}_\textbf{SGCMINV}||_\infty$.\\
$$Table1.$$
$$
\begin{tabular}{|c|c|c|c|}
  \hline
   %after \\: \hline or \cline{col1-col2} \cline{col3-col4} ...

  \multirow{2}{*}{} & \multicolumn{3}{c|}{n} \\ \cline{2-4}
   & 50 & 100 & 500 \\ \hline
  $\varepsilon=||C^{-1}_\textbf{exact}-C^{-1}_\textbf{SGCMINV}||_\infty$ & $1.1631\times 10^{-9}$ & $1.1215\times 10^{-9}$ & $1.6078\times 10^{-9}$ \\ \hline
  \multirow{2}{*}{CPU time(s)} & 0.421(using inverse function in Maple 13.0) &3.448 & 336.338 \\ \cline{2-4}
   &0.109(Using our Algorithm)  &0.609  & 33.899 \\ \hline 
  \end{tabular}
$$
%\textbf{Acknowledgements} The author likes to thank Prof. Dr M.E.A. El-Mikkawy and referees for several comments and suggestions.

\section{CONCLUSIONS}

In this work new symbolic computational algorithms have been developed for computing the determinant and inverse of a general comrade matrix. The algorithms are reliable, computationally efficient and remove the cases where the numeric algorithms are fail.

\end{document}